\documentstyle[prb,aps]{revtex}
\begin{document}
\bibliographystyle{prsty}
\draft
\title{Elementary excitations in the $\Delta$-chain}
\author{Tota Nakamura and Kenn Kubo}
\address{Institute of Physics, University of Tsukuba,
         Tsukuba, Ibaraki 305, Japan }
\date{\today}
\maketitle
\begin {abstract}
We clarify elementary excitations in the $\Delta$-chain.
They are found to be  `kink'-`antikink' type domain wall excitations
to the dimer singlet ground state.
The characters of a kink and an antikink
are quite different in this system: a kink has no excitation energy and is
localized, while an antikink has a finite excitation energy and propagates.
The excitation energy of a kink-antikink pair consists of a finite energy gap
and a kinetic energy due to the free motion of the antikink.
Variational wave functions for an antikink
are studied to clarify its propagating states.
All the numerical results are explained consistently based on this picture.
At finite temperatures, thermally excited antikinks
are moving in regions bounded by localized kinks.
The origin of the low-temperature peak in  the specific heat
reported previously is explained and the peak position
in the thermodynamic limit is estimated.
\end  {abstract}

\section{Introduction}
\label{sec:intro}
Recently, much interest is focused on systems whose classical
(mean-field) ground states exhibit an infinite number of local continuous
degeneracies due to frustration(``floppy" systems).
In such a system, construction of the linear spin-wave theory based on
one of the classical ground states leads to at least one spin-wave
mode whose frequency vanishes for all wave vectors(zero-energy mode).
The zero-energy mode corresponds to local deformation of a spin configuration
that does not raise the energy.
The set of the ground state configurations is a manifold with dimensions
proportional to the system size.
The classical ground state thus may be considered as disordered. %
We can find examples of floppy spin systems in all dimensions.\cite{ramirez94}
It is an interesting and a challenging problem to investigate how the quantum
effects manifest themselves in such spin systems.
The central issue is whether quantum effects lift
the ground state degeneracy and select some long-range order.
Also the low-energy excitation spectrum is of interest,
since it may lead to peculiar thermodynamic properties.

A typical example of such systems is the antiferromagnetic Heisenberg
(AFH) model on the {\it kagom\'e} lattice.
Intensive studies
\cite{elser89,zeng-e90,chalker-ws92,keren94,reimers-b93,harris-kb92,%
chubukov92,sachdev92,leung-e93,elstner-y94,zeng-e95,nakamura-m95}
on this system have been inspired by the experiments on the $^3$He layer
adsorbed on graphite \cite{greywall89,franco86}
and also on the compound SrCr$_8$Ga$_4$O$_{19}$.\cite{fioriani-dts85}
What mainly have been concerned with are
the existence of a double peak in the specific heat at low temperatures
and
whether a magnetic order is realized in the ground state or not.
Several approximate analyses \cite{harris-kb92,chubukov92,sachdev92}
have been done but they are still far from giving a common understanding
on the ground state property.
Numerical studies on finite systems support the existence of the double
peak.\cite{elser89,zeng-e90,zeng-e95,nakamura-m95}

In this paper we study a simple one-dimensional floppy
system called the $\Delta$-chain.\cite{hamada-knn88,doucot-k89}
We consider that this model shares general features of quantum floppy
systems, and an understanding of its properties might give some insight into
that of the {\it kagom\'e} antiferromagnet.
The ground state of this system is exactly known to be a dimer
state.\cite{monti-s91,monti-s92}
One of the authors(K. K.) examined the low-lying
excitations and the specific heat in a previous paper.\cite{kubok93}
The numerical diagonalization study of small clusters (up to 20 spins)
exhibited that the low-lying excitation modes in the periodic
chains are almost dispersionless.
The first and the second lowest mode were revealed to converge
to dispersionless modes with the same energy in the
thermodynamic limit.
The specific heat was shown to have a double peak in common with
the {\it kagom\'e} antiferromagnet.
This double peak structure was also observed
recently by a Monte Carlo method \cite{nakamura-s95}
and by a recursion method.\cite{otsuka95}

The dispersionless aspect of excitations may be considered to imply immediately
their localization and hence very weak size-dependence of various quantities.
The energy gap obtained by numerical diagonalizations,\cite{kubok93}
however, does exhibit fairly large size-dependence.
Also a broad bump of the spin correlation was observed between the most distant
spin pairs in the lowest triplet excitations.\cite{kubokunp}
These results suggest that the excited states are not localized.
We solve this puzzle and clarify the character of
low-lying excited states in the following sections by mainly employing
numerical diagonalization of finite size systems.
Elementary excitations are revealed to be
`kink'-`antikink' type domain walls created in the singlet dimer ground state.
An antikink is shown to move freely in a region bounded by kinks at both ends.
The kinetic energy of an antikink leads to the size dependence
of the energy gap.
On the other hand, a kink is localized and gives the dispersionless
property to a kink-antikink pair excitation mode.

Properties of an isolated kink and an antikink, and also interactions between
them will be discussed in Sec. \ref{sec:elem}.
In Sec. \ref{sec:num}, we show that the size dependence of the
energy gap in the periodic chain agrees with that of the
kinetic energy of an antikink.
The specific heat due to kink-antikink excitations is discussed in
\ref{sec:spec},
and the position of the low-temperature peak in the thermodynamic limit is
estimated.
Conclusion is given in Sec. \ref{sec:conc}.

\section{Elementary excitations}
\label{sec:elem}
\subsection{Introduction of a kink and an antikink}
The $\Delta$-chain is described by the following Hamiltonian.
\begin{equation}
  H=\sum_{i=1}^N h_i,
  \label{eq:hamil}
\end  {equation}
where
\begin{equation}
  h_i=\mbox{\boldmath $S$}_{2i-1}\cdot\mbox{\boldmath $S$}_{2i  }
     +\mbox{\boldmath $S$}_{2i  }\cdot\mbox{\boldmath $S$}_{2i+1}
     +\mbox{\boldmath $S$}_{2i-1}\cdot\mbox{\boldmath $S$}_{2i+1}.
  \label{eq:hi}
\end  {equation}
$\mbox{\boldmath $S$}_i$ is the spin with size $1/2$ at the site $i$
and $N$ denotes the number of triangles in the chain.

We first explain the ground state of this system and then forward to the
explanation of elementary excitations.
Under periodic boundary conditions, the ground state is the perfect singlet
dimer state, since the ground state of the local Hamiltonian $h_i$ is
realized by pairing any two of three spins into the singlet state.
The two-fold degenerate ground states
are written as
\begin{eqnarray}
  \psi_{\rm g}^{\rm L}&=&[1, 2]\otimes[3, 4]\otimes[5, 6]
                              \otimes\cdots\otimes[2N-1, 2N]\nonumber \\
  \psi_{\rm g}^{\rm R}&=&[2, 3]\otimes[4, 5]\otimes[6, 7]
                              \otimes\cdots\otimes[2N, 1].
  \label{eq:psig}
\end  {eqnarray}
Here $[i, j]$ denotes the singlet dimer of the spins at the site $i$ and $j$,
i.e., $[i, j]=(\alpha_{i}\beta_{j}-\beta_{i}\alpha_{j})/\sqrt{2}$,
where $\alpha_{i}$($\beta_{i}$) is the state with $S_{i}^{z}=1/2(-1/2)$.
They are schematically depicted in Fig. \ref{fig:scheme} (a) and (b).
We call a dimer located on the left(right) side of a triangle an L-dimer
(R-dimer) and
hence the state $\psi_{\rm g}^{\rm L}$($\psi_{\rm g}^{\rm R}$) an L-dimer
(R-dimer) state.
These two states are linearly independent but not
orthogonal to each other for finite $N$.
They become orthogonal in the limit where $N\rightarrow \infty$.
The existence of the excitation energy gap above the ground state was
rigorously shown.\cite{monti-s91,monti-s92}

Under open boundary conditions, the ground state of a system with $N$
triangles is highly degenerate
since a ground state consists of $N$ singlet dimers and one free spin.
One of the ground states is shown in Fig. \ref{fig:scheme}(c).
A dimer configuration is uniquely determined by fixing the position of
the free spin.
It is seen in the figure that the free spin plays a role of a domain wall
between an L-dimer state and an R-dimer state.
The number of possible positions for a free spin in the ground state
is equivalent to the number of sites, and therefore
$2N+1$ different dimer configurations can be considered.
However, those $2N+1$ states are linearly dependent, since only two out of
three dimer configurations in a triangle are linearly independent.
The dimensions of the ground state reduce to
$2\times (N+1)$;\cite{monti-s91,monti-s92}
only one positional freedom per triangle is allowed.

Let us consider a simple trial excited state in a periodic chain
constructed by changing one dimer singlet
in the ground state into a triplet state.
Then the expectation value of the excitation energy  $\Delta E$ is identical
to the singlet-triplet gap; $\Delta E=1$.
Since the energy gap is known as $\Delta E\sim 0.22$
in the thermodynamic limit,\cite{kubok93}
the lowest excited state is not such a simple state.
In fact, the above state is not an eigenstate and the two spins coupled to a
triplet are separated when the Hamiltonian is operated.
The average excitation energy may decrease if the two spins are
apart
as depicted in Fig. \ref{fig:scheme} (d).
Each of two spins is regarded as a domain wall
between an L-dimer and an R-dimer state.
We show in the following that isolated domain walls are the elementary
excitations in this system.

A domain wall which has L-dimers on the left side and R-dimers on the right
has a singlet dimer in its own triangle, and thus
is the ground state of the local Hamiltonian $h_i$.
We call this type of a domain wall a `kink'.
A free spin appearing in the ground state of the open $\Delta$-chain is
a domain wall of this type.
Therefore the excitation energy of a kink is null.
There is ambiguity in the definition of the position of a kink,
since ground states with different positions of the free spin
are not orthogonal to each other.
Let us consider a kink as a state where a free spin is located at the top of a
triangle as depicted in Fig. \ref{fig:scheme} (c) for a convention.
Then the overlap between the states with an up(or a down)-spin at the  $i$th
and the $j$th($i\ne j$) triangle is $-(-2)^{-|i-j|}$.
We treat a kink as a localized object since the state with a kink alone
is an eigenfunction of the system,
although its position is not a good quantum
number due to the non-orthogonality.

We call another type of a domain wall an `antikink',
which has R-dimers on the left side and L-dimers on the right.
The state with an antikink at the $i$th triangle is not a ground state of
$h_i$ as the triangle can not accommodate a singlet dimer.
A finite energy is necessary to excite an antikink.
It spreads out to other triangles and propagates among them.
The energy of an antikink is lowered by this motion,
which will be discussed in the next subsection.

Both a kink and an antikink have a spin $1/2$ as a whole, and
must appear alternatively.
Elementary excitations in the Majumdar-Ghosh model
is known as propagating domain walls.\cite{majumdar-g69,shastry-s81}
They are also kink-antikink type domain walls.
The physical properties of a kink and an antikink are same in this
model, since they are transformed to each other by symmetry operations.
On the other hand, the properties of a kink and those of an antikink are
quite different in the $\Delta$-chain:
kinks are localized and antikinks move about
in a region bounded by two localized kinks.
In the following,
we study first an isolated antikink and then the interaction between a kink and
an antikink.

\subsection{An isolated antikink}
An antikink is necessarily accompanied by a kink in a periodic chain and it is
not easy to extract the properties of an isolated  antikink
because the translational symmetry mixes the states with different positions
of a kink and an antikink.
We can solve this difficulty by considering open systems as shown in
Fig. \ref{fig:openscheme} (a), which we call `Open-A' systems.
In this subsection we treat only Open-A systems.

In this system,
two additional edge bonds have strong tendency to form dimer singlet states
and they force the left(right)-hand side of the system
to be in the R(L)-dimer state.
Then inconsistency of the dimer configurations results in an antikink
to appear in the middle of the chain.
Thus we expect only an antikink to exist
in the ground state of an Open-A system.
We confirm this speculation and clarify the
properties of an isolated antikink through numerical and variational analyses.
The energy of an antikink $\delta E$ is obtained from the total energy $E$
of the system with $N$ triangles as $\delta E = E +3(N+2)/4$.
Figure \ref{fig:open-acorsz} shows the nearest-neighbor spin correlation
$\langle \mbox{\boldmath $S$}_i\cdot \mbox{\boldmath $S$}_{i+1} \rangle$
and the local magnetization $\langle S_{i}^{z} \rangle$
in the lowest four states of an `Open-A' system with $S^{z}=1/2$.
We consider the ground state first.
It is seen that the spin correlations at the both edge bonds are very close to
the value of a dimer singlet state $-3/4$, while those of
the bonds next to the edges almost vanish.
These features show that the both edges are almost in the perfect dimer states.
We notice that the state close to the left(right) edge
is approximately the R(L)-dimer state.
As the position of a bond approaches the center,
the singlet coupling becomes weak and finally the phase changes at the center.
Since the L-dimers are located in the right-hand side of the domain wall,
we may conclude that there is an antikink.
The local magnetization oscillates extended to the whole system,
and its amplitude has a broad maximum in the middle of the system.
The system-size dependence of the domain wall profile in the ground state is
shown in Fig. \ref{fig:open-adom}.
We plotted as the profile the deviation of the nearest-neighbor correlation
from that in the R(L)-dimer state on the left(right) half of the system.
That is,
$0.75+\langle \mbox{\boldmath $S$}_i\cdot \mbox{\boldmath $S$}_{i+1} \rangle$
for odd positive or even negative $i$  and
$0-\langle \mbox{\boldmath $S$}_i\cdot \mbox{\boldmath $S$}_{i+1} \rangle$
for even positive or odd negative $i$.
We notice in this figure
that the profile of the domain wall is extended to almost whole region
of the chain.
Results of the profile and the magnetization in the ground state of Open-A
systems imply that the antikink is in a state
extended to the entire region of the system with large amplitude in the
middle.

In the first excited state,  the profile of the magnetization has two bellies
at about a quarter of the system size from both edges and a node
at the center.
If we assume that the local magnetization reflects
the probability of existence of an antikink, the above result suggests that
the wave function of an antikink is something like a sine-function with a node
at the center.
The spin correlation profile is very flat with its value $\sim -0.35$ in the
middle region.
The correlation at a bond is averaged to be $-3/8$ if the probability for
an antikink to be at the left of the bond is equal to that at the right.
So the result is consistent with a wave function of an antikink
which has a node in the center and two bellies.
We can also speculate that the wave functions of
the second and the third excited states have two and three nodes, respectively.
The flat region in the spin correlation
in the third excited state suggests existence of a node at the center.
We can not, however, extract detailed properties of the wave functions
for these excited states from the present numerical data alone.
We see in the following that our speculation is correct by
comparing the diagonalization results with those by a simple trial functions.

Numerical results of Open-A systems suggest that an antikink behaves like a
free particle propagating in the whole system,
since the data are consistent with that the wave function
of an antikink is a sine function in terms of the position.
In order to check the validity of this picture we have made a variational
analysis.
We take the state where an antikink consists of one free spin
as the first trial function.
Then the variational basis is the set of $ \psi_i^{(1)}$
which consists of $(N+1)$ dimer singlet
pairs and the one free spin located at the top of the $i$th triangle
as depicted in Fig. \ref{fig:varbase}(a).
The dimension of the basis is $N+2$ since the free spin may occupy the sites at
the edges.
These basis functions are not orthogonal to each other and satisfy the
following relations.
\begin{eqnarray}
 \langle \psi_i^{(1)}|\psi_j^{(1)}\rangle&=&\left (-\frac{1}{2}\right )^{|i-j|}
\\
 \langle \psi_i^{(1)}|H|\psi_j^{(1)}\rangle&=&-\frac{3}{4}
   [(N+2)\langle \psi_i^{(1)}|\psi_j^{(1)}\rangle-\delta_{ij}]
\end  {eqnarray}
Here, $\delta_{ij}$ is the Kronecker delta.
The trial function $\Psi_{\rm var}^{(1)}$, which is assumed as
\begin{equation}
 \Psi_{\rm var}^{(1)}\equiv \sum_{i=0}^{N+1} C_i\psi_i^{(1)},
 \label{eq:varf}
\end  {equation}
is determined as an eigenfunction of the matrix
$\langle \psi_i^{(1)}|\psi_j^{(1)}\rangle$ ($0\leq i,j \leq N+1$).
In the limit of infinite $N$ we can neglect the boundary effect, and
the eigenfunction is  easily obtained as
$C_i\propto (-1)^i \sin {qi}$  with the excitation energy
$\delta E_{\rm 1-spin}=5/4-\cos q$.
For a finite $N$, we have solved the eigenvalue problem numerically
and have found that the $k$th state($k= 1, 2, \cdots$) and
its antikink energy is well approximated by
\begin{eqnarray}
  C_i&\propto &(-1)^i \sin \frac{\pi k (i+3/2)}{N+4}
\end{eqnarray}
and
\begin{eqnarray}
  \label{eq:1spinvar}
  \delta E_{\rm 1-spin}&=&\frac{5}{4}-\cos \frac{\pi k}{N+4}.
\end  {eqnarray}
The variational energy gap converges to
$\delta E_{\rm 1-spin}= 0.25$ as $N\to\infty$.
The fact that the true energy gap converges to about 0.22 implies
that the above choice
of the variational basis is too simple to describe an antikink.
We have to take into account of the fact that
a free spin spreads out to neighboring triangles
destroying singlet dimers nearby.
As an improved trial function, we have employed $\Psi_{\rm var}^{(5)}$
which is given by Eq. (\ref{eq:varf}) with $\psi_i^{(1)}$ replaced with
$\psi_i^{(5)}$, which consists of $(N-1)$ dimer singlet pairs and
the ground state of the cluster
with five spins as depicted in Fig. \ref{fig:varbase} (b).
We numerically diagonalized the effective Hamiltonian
for $\Psi_{\rm var}^{(5)}$.
The wave functions for the lowest three levels
obtained numerically for the system with 98 triangles
are shown in Fig. \ref{fig:varwave}.
The wave functions ($C_{i}\times (-1)^{i}$) look like sine functions
and is consistent with the numerical
results of the local magnetization shown in Fig. \ref{fig:open-acorsz}.
There is no qualitative difference between the 1-spin variational
wave functions and the 5-spin ones.

The variational energy by $\Psi_{\rm var}^{(5)}$ is lower
than that by $\Psi_{\rm var}^{(1)}$ for low-lying states($k\ll N$) and
leads to the energy gap $\Delta E_{\rm 5-spin}\sim 0.23$
in the limit $N\to\infty$.
The 5-spin variational energy, however, rapidly increases
as $k$ approaches $N$ and exceeds the 1-spin variational energy,
which implies that the 5-spin wave function is not a good approximation
in the high momentum region.
This is because we only took into account the
{\it ground state} of the 5-spin Hamiltonian.
In the high momentum region the 1-spin variational function seems to be a
better approximation than the 5-spin one.
The expectation value of the local magnetization in the 1-spin variational
wave function $\Psi_{\rm var}^{(1)}$ is compared with the exact results
for the lowest three states in Fig. \ref{fig:varsiz}.
Overall behaviors of both results agree very well not only for the lowest
state but for the excited states.
This fact supports that the low energy states of the
system are those with a freely propagating antikink.
The amplitude of the oscillations of the variational results is larger
than the exact ones.
This reflects that the real antikink is not a single spin
as supposed in the 1-spin variational function but
extended to several triangles and thus the magnetization is smeared.
The size dependence of the variational energy is shown
in Fig. \ref{fig:altogether}
and is compared with the diagonalization results.
All the data show almost linear dependence on $N^{-2}$.
It is seen that the slope of the results of the 5-spin trial function
agrees very well with the diagonalization results.
Another important fact shown in this figure is that
the behavior of the eigen energy of the Open-A system
agrees with that of the excitation energy of the periodic system.
A low energy excitation of the periodic system is considered to be
a kink-antikink pair excitation but
the above result implies that the excitation energy
in the periodic chain is mainly determined by a freely moving antikink.

\subsection{kink-antikink interaction}

In this subsection, we investigate the wave function of a kink and
the interaction between a kink and an antikink.
For this purpose we study the excited states of finite clusters as depicted in
Fig. \ref{fig:openscheme} (b), which we call `Open-B' systems.
The L-dimer state is the unique ground state of this Open-B system.
Hence low energy excitations in this system necessarily involve
a kink and an antikink and
the open boundary conditions make each of them visible as shown below.
We first consider the first excited state.
This is a triplet state and we show the nearest-neighbor spin correlation
$\langle \mbox{\boldmath $S$}_i\cdot \mbox{\boldmath $S$}_{i+1}\rangle$ and
the local magnetization $\langle S^z_i\rangle$ in the state with $S^z=1$
in Fig. \ref{fig:open-bedge}(a).
The numerical results show that the leftmost bond forms a complete
triplet state for all system sizes, i.e.,
$\langle \mbox{\boldmath $S$}_1\cdot \mbox{\boldmath $S$}_{2}\rangle =1/4$
(note that $\mbox{\boldmath $S$}_1\cdot \mbox{\boldmath $S$}_{2}$
commutes with the Hamiltonian).
The data for the three leftmost spins approach the limiting values
when $N\to \infty$ as:
$\langle \mbox{\boldmath $S$}_2\cdot \mbox{\boldmath $S$}_{3}\rangle \to -1/2$,
$\langle S_1^z\rangle=\langle S_2^z\rangle\to 1/3$ and
$\langle S_3^z\rangle\to -1/6$.
In fact the difference from the limiting values is almost negligible
in the data of the system with $N=11$ (not shown in the figure).
This result implies that the total wave function reduces to
a direct product of the wave function of these three spins given by the
following equation
and that of the remaining $(2N-1)$ spins in the limit $N\to \infty$.
\begin{equation}
 \psi_{\rm kink}=[\alpha_1(\alpha_2\beta_3-\beta_2\alpha_3)
                 +\alpha_2(\alpha_1\beta_3-\beta_1\alpha_3)
                 ]/\sqrt{6}
 \label{eq:kink}
\end  {equation}
Therefore a kink is localized in the leftmost triangle
though it is a linear combination of the wave function
with two different positions of a free spin.
An antikink is distributed in the rest part of the system
which corresponds to an Open-A system with $N-2$ triangles.
It also feels a magnetic field at the 4th and 5th site
caused by a magnetization $\langle S_3\rangle\sim -1/6$
of the kink at the third site.
Otherwise the antikink is freely propagating in this region.
The domain wall profile shown in Fig. \ref{fig:open-bedge}(a) is
fairly well reproduced by that of the Open-A system with 6 triangles
and with a magnetic field $-1/6$ at the site 1and 2.
Almost complete reproduction of the profile is made by
adjusting the magnetic field to be $-0.225$,
which implies that the antikink feels a larger effective field
due to the overlap of the wavefunction with that of the kink for finite $N$.

We have examined higher excited states up to the fourth
in the subspace of $S^{z}=1$. (Not shown in the figure.)
We found that a kink is located in the $n$th triangle
in the $n$th excited state and
the left side of that triangle is in the L-dimer state.
Therefore the wave function of the $n$th state is
nothing but a direct product of the wave function of the L-dimer
state of $n-1$ triangles and that of the first excited state of
a smaller Open-B system with $N-(n-1)$ triangles.

Excitations corresponding to excited states of the Open-A systems
can be obtained by restricting the phase space
to that where the edge bond forms a triplet state
(shown in the Fig. \ref{fig:open-bedge}(b),(c)).
This restriction ensures the existence of a kink at the leftmost triangle
and then the domain for an antikink is same with the first excited state.

It is seen that the results at the leftmost triangle
scarcely vary among four excitations in Fig. \ref{fig:open-bedge}.
This implies that the state of a kink is almost same in these three states
and is localized within the leftmost triangle.
The results on the right of the leftmost triangle agree very well with
those in the `Open-A' system shown in Fig. \ref{fig:open-acorsz}.
Only a small difference is seen at the center of the chain where the
domain wall profile is not flat for the Open-B system.
This difference shows that the antikink is pulled to the left
due to the effective exchange interaction with the kink.
Since the leftmost triangle is occupied by a kink,
the region for the motion of an antikink in the
system with $N$ triangles is same with that in a Open-A system
with $N-2$ triangles.
The interaction with a kink attracts the wave function of an antikink,
and as a result, an antikink feels a larger region for its motion.
The wave function in the $S=1$ subspace roughly corresponds
to that in the Open-A system with $N-1$ triangles.
We discuss this problem in detail in the next section.

We have also obtained the singlet excited states in the subspace
where the edge bond forms a triplet state.
The result of the lowest excited state is shown in Fig. \ref{fig:open-bs01}(a)
together with that in the triplet state.
We find that the results at the leftmost triangle is almost same with
those in the corresponding triplet states.
On the other hand, the results on the right side of the system are different
from those in the the triplet states.
That is, the domain wall profile is shifted to the right
by nearly one triangle compared to that in the triplet states.
This is the effect of the exchange
interaction with the kink, which acts as a repulsion in the singlet state.
The profile of the singlet state shown in Fig.\ref{fig:open-bs01}(a)
is fairly well reproduced by that in the ground state
of the Open-A system with 6 triangles under
the magnetic field $1/6$ at the site 1 and 2.
More precise reproduction of the profile can be made by
adjusting the magnetic field to be $0.320$.
The fact that the antikink feels stronger effective field
in the singlet state than in the triplet one seems to imply that
the kink is extended to outside of the leftmost triangle
for this size of the system.
The antikink in the singlet state is repelled stronger
by the kink than that in the triplet state
because the state must be orthogonal to the ground state.
This effect might be the cause of the stronger effective field,
and as a result, it is nearly prohibited for the antikink
to occupy the site 4 in the singlet state.
The wave function of an antikink in the $S=0$ subspace
roughly corresponds to that in the Open-A system with $N-3$ triangles.
In Fig. \ref{fig:open-bs01}(a),
we compare the domain wall profiles of the lowest
triplet and the singlet excitations.
We see that the center of the profile is different nearly by
one triangle between two cases,
which implies that the effective length
of the region where the antikink can move is different by about two triangles.
The profiles of the the ground state of the Open-A system with $N$ triangles
under the magnetic field $-0.225$,
that of the $S=0$ excited state of the Open-B system with $N+3$ triangles,
and the mirror image of the profile of the $S=1$ excited state
of the Open-B system with $N+1$ triangles agree surprisingly well
in the central region of the antikink(Fig \ref{fig:open-bs01}(b)).
We may conclude by this evidence that the wave functions of an antikink
in these three states are approximately equal.

We have seen above that a kink is localized
and an antikink is moving in the region
bounded by the kink and by the open boundary in the low energy excited states
of the Open-B systems.
The interaction between a kink and an
antikink is approximated by a hard-core repulsion on the same triangle and a
ferromagnetic exchange interaction on the nearest neighbor triangles.
The exchange constant is approximated by $-1/3$ in the limit of infinite $N$.
Otherwise an antikink can be considered as a free particle.
A kink is pushed to the left edge by an antikink as to lower
its kinetic energy in Open-B systems.

We have studied the interaction between a kink and an antikink
when only one antikink is on the right side of the kink.
In periodic systems and/or higher excitations,
an antikinks exists also on the left side of a kink.
We expect that a kink will be localized even
when antikinks exist at both sides.
Its wave function, however, would be different from that studied above
which is asymmetric with respect to the inversion.
So we have obtained the lowest excitation with $S^{z}=3/2$
of the Open-A system with 17 spins(7 triangles) to
investigate excitations with a kink and two antikinks.
The system size, however, seems to be too small to study a
detailed nature of the kink-antikink interaction
from the obtained domain wall profile and the local magnetization.
We could not obtain the corresponding state in the system
with 21 spins by the Lanczos method because of a poor convergence of the
iteration.
We conjecture that
a ferromagnetic exchange interaction acts between a kink and an antikink
when they are at the nearest neighbor triangles.
The ferromagnetic nature of the interaction is confirmed
from the result of the periodic systems
where the lowest excited state is triplet.
More study is necessary to determine the magnitude of the exchange coupling.

\section{Size dependence of the energy gap}
\label{sec:num}
In the previous section, we focused on a kink and an antikink
as elementary excitations and clarified their individual characteristics
and the interaction between them.
We found that the excitation energy of the system
is determined mainly by the excited antikink.
The antikink is described as a particle propagating freely
in a region bounded by kinks and/or boundaries.
If the region of its motion is bounded by a kink,
it feels an exchange field which modifies its wave function.
The expectation value of the exchange interaction is usually very small,
since the probability for a kink and an antikink to be
at the nearest neighbor triangles is very small.
The effect of the interaction appears rather as a change
in the effective length of the region of motion for the antikink.
We therefore postulate that the excitation energy of the system is
given by the energy of a freely-propagating antikink as a first approximation.
Then the excitation energy may be written as
\begin{equation}
 \delta E=\epsilon_{\rm a}
         =\epsilon_{0} + \frac{1}{m}(1-\cos \frac{\pi k}{N'+1}).
\label{eq:deltaE}
\end  {equation}
Here $\epsilon_{\rm a}$ denotes the energy of an antikink, $m$ its mass,
$k-1$ stands for the number of nodes in the wave function and
$N'$ the effective length of the region where the antikink propagates.
The creation energy $\epsilon_{0}$ of an antikink is equal to the excitation
energy gap of the $\Delta$-chain in the thermodynamic limit.
Of course, the cosine form in Eq. (\ref{eq:deltaE}) is an approximate
form only valid in the region where $\pi k/(N'+1)$ is small compared to unity.
We treat only such cases.
In this section, we show that
the excitation energies in the periodic systems as well as
those in the open systems considered in the previous section
are well described by Eq. (\ref{eq:deltaE}).
Before comparing Eq. (\ref{eq:deltaE}) with numerical results in detail,
we examine the size dependence of the excitation energy in the periodic chain.

In Fig. \ref{fig:periodicgap} we show the size dependence of
the excitation energies $\delta E$ in the subspace with zero total momentum
classified by the total spin and the parity
in the periodic systems with up to 13 triangles.
The energy levels up to the 7th excitation
in each subspace have been obtained by the Lanczos method.
In the figures for the $S=0$ and the $S=1$ subspace, we find that $\delta E$
for the lowest levels converge to the same value $\sim 0.22$
in the thermodynamic limit.
In the subspace with even parity, $\delta E$
of the second level also seems to converge to the same value.
These levels are considered to be states with a single pair of
a kink and an antikink.
The effect of the exchange interaction between a kink and an antikink
vanishes as $N\to \infty$ and so $\delta E$ for $S=1$ and
$S=0$ converge to the same value corresponding to $\epsilon_0$.
On the other hand, the lowest $\delta E$ in the $S=2$ and $S=3$ subspaces
seem to converge to $2\epsilon_0$ and $3\epsilon_0$, respectively.
The results are consistent with the picture that
the lowest excitation in the $S=2$($S=3$) subspace is a state
with two(three) pairs of a kink and an antikink.
It is still not conclusive whether they are precisely twice or three times,
since the systems we have treated are not large enough
to extrapolate such high excitations.
We also observe that $\delta E$ of higher excitations in $S=0$ and $S=1$
subspaces seems to converge to $2\epsilon_0$.
These states are also regarded as states
with two pairs of a kink and an antikink.
The results mentioned above indicate that
kinks and antikinks behave as independent $S=1/2$ particles
when they are far apart.
We further confirm the above picture by comparing the size dependence of
all the systems we have considered in this paper.
We examine here $\delta E$ of the ground state of the Open-A systems,
the first excited state of the Open-B systems with $S=0$ and that with $S=1$
and
the lowest three excited states of the periodic systems with $S=0$ and those
with $S=1$ in the subspace with zero total momentum.
We assume that
Eq. (\ref{eq:deltaE}) with $k=1$ expresses
$\delta E$ of the ground state of the Open-A systems,
the first excited state of the Open-B system with $S=0$ and that with $S=1$.
The first, the second, and the third excited states of the periodic systems
with $S=0$ and $S=1$ are also assumed to be expressed by
Eq. (\ref{eq:deltaE}) with $k = 1, 2,$ and $3$, respectively.
The parity of these states are consistent with this assignment
(parity = even for $k =1$ and 3; parity = odd for $k=2$).
In Fig. \ref{fig:cosineplot}, we show the plot of $\delta E$ versus
$1-\cos(\pi k/(N'+1))$,
where $N'=N+\Delta N$ is chosen so that all the data fall onto one curve.
$\Delta N$ is a correction to $N$ induced by the boundary effect and/or
the interaction with kinks.
It is remarkable that all data lie on a curve.
This is a clear evidence for that the excitation energy is mainly
contributed by an antikink.
For the Open-A systems we take $\Delta N=3$ for a convention.
Then the best choice of $\delta N$ for other cases
are shown in Table I.

Let us try to understand the above results in an intuitive way.
In Open-B systems, a kink occupies the leftmost triangle.
Then the region allowed for an antikink to move is assumed to be equivalent to
an Open-A system with less triangles by two.
The attractive interaction elongates the effective region for $S=1$ state,
and repulsive interaction and the requirement of the orthogonality to the
ground state shortens it for $S=0$.
We see from the numerical results (Table I)
that the effect of the interaction amounts
to $1.3 (-0.2)$ triangles for $S=1(0)$.
It amounts to $-0.2 (-3)$ triangles for $k=1$ and $S=1(0)$ states
in the periodic systems,
since the allowed region for an antikink is equivalent to
an Open-A system with less triangles by three.
The difference of the effect of the interaction and/or the orthogonality
between the Open-B systems and the periodic systems
is not clarified yet at present.
The least-square extrapolation by using all the data in the figure
gives the creation energy $\epsilon_0$ in the limit $N \to \infty$
and the mass of an antikink $m$.
They are
\begin{equation}
  \epsilon_{0}=0.215 \label{eq:ep}\\
\end{equation}
and
\begin{equation}
  m =1.21. \label{eq:mass}
\end  {equation}

\section{The Low-temperature specific heat}
\label{sec:spec}
In this section, we show that the elementary excitations discussed
in the previous sections generate the low-temperature peak in the specific
heat.
First we calculate the specific heat of a finite system
by numerical diagonalization, and examine whether the result can be
reproduced by the elementary excitation with the spectrum given by
Eq. (\ref{eq:deltaE}) with the parameters (\ref{eq:ep}) and (\ref{eq:mass}).
We consider a periodic system with 8 triangles ($N=8$).
All the $2^{16}$ states can be diagonalized numerically, and
thus the exact specific heat is obtained as shown in Fig. \ref{fig:spec}.
Now we calculate the specific heat due to kink-antikink excitations.
We take only low-lying states into account, i.e.,
the ground state (doubly degenerate), all the
states with one pair of a kink and an antikink,
and some states with two pairs.
Among the states with two kink-antikink pairs,
we take those where the distance between two kinks are largest.
We consider that these are  more important than others since
the region for the motion of two antikinks are comparatively large.
We checked that these states
contribute to the low-temperature behavior of the specific heat
by comparing the results that we took all the configurations into account.

The expectation value of energy  are written approximately as
\begin{equation}
 \langle E\rangle _{\beta} = E_{\rm g}+
\frac{\displaystyle{
 8\sum_{S_1=0}^1 \sum_{k_1=1}^7 A_{S_1}
   E_{S_1}^{(8)}\exp[-\beta E_{S_1}^{(8)}]
 +4\sum_{S_1, S_2=0}^1 \sum_{k_1, k_2=1}^3 A_{S_1}A_{S_2}
  (E_{S_1}^{(4)}+E_{S_2}^{(4)})\exp[-\beta (E_{S_1}^{(4)}+E_{S_2}^{(4)})]
                                  }}
                                  {\displaystyle{
  2
 +8\sum_{S_1=0}^1 \sum_{k_1=1}^7 A_{S_1}
   \exp[-\beta E_{S_1}^{(8)}]
 +4\sum_{S_1, S_2=0}^1 \sum_{k_1, k_2=1}^3 A_{S_1}A_{S_2}
   \exp[-\beta (E_{S_1}^{(4)}+E_{S_2}^{(4)})]
                                  }}, \label{eq:Eav}
\end  {equation}
with
\begin{eqnarray}
 E_{\rm g}    &\equiv& -\frac{3}{4}N \\
 E_{S_i}^{(N)}&\equiv& \epsilon_0+\frac{1}{m}\cos \frac{\pi k_i}{N+1}
       ~~~\mbox{\rm for $S_i=1$}  \label{eq:ET} \\
 E_{S_i}^{(N)}&\equiv& \epsilon_0+\frac{1}{m}\cos \frac{\pi k_i}{N-2}
       ~~~\mbox{\rm for $S_i=0$}. \label{eq:ES}
\end  {eqnarray}
$\epsilon_0$ and $m$ are given by Eqs. (\ref{eq:ep}) and (\ref{eq:mass}).
$S_i$ denotes the total spin value of the $i$th pair of a kink and an antikink.
Here we have treated the states with two kink-antikink pairs
as two independent pairs for simplicity.
Also we have used integer values for $\Delta N$ in Eqs. (\ref{eq:ET})
and (\ref{eq:ES}).
The spin multiplicity is denoted by $A_S$,
namely $A_S=3$ for $S=1$ and $A_S=1$ for $S=0$.
The specific heat is calculated by differentiating
this equation with respect to the temperature.
The result coincides with the exact specific heat at low temperatures
as shown in Fig. \ref{fig:spec}.
The peak position and the low-temperature side
of the peak are correctly reproduced.
The peak height is a little smaller than the exact value.
The approximate result gives much smaller value than
the exact one at higher temperatures
because of the negligence of higher excited states.
We may conclude that the low-temperature peak is caused
by the low-lying kink-antikink excitations whose spectrum
is described by Eq. (\ref{eq:deltaE}).

Next we calculate the  specific heat in the thermodynamic limit.
We consider the system at low temperatures
where the thermal excitations are described by kinks and antikinks.
Let us first consider the case where the mass $m$ is infinite.
In this case, antikinks are localized with the excitation
energy $\epsilon_{0}$.
This system is thermodynamically equivalent to that
where both kinks and antikinks have the excitation energy $\epsilon_{0}/2$,
and it shows a Schottky-type specific heat with an excitation energy
$\epsilon=\epsilon_{0}/2$.
If $m$ is finite, the excitation energy of an antikink
$\epsilon_{\rm a}$ depends on the length $L$
of the region of the motion and on the level $k$ as
$\epsilon_0+(1/m)\cos(\pi k/L)$.
In order to obtain thermodynamic quantities, we must
average them over all states with possible values of $L$ and $k$.
Instead of doing this summation, we calculate the specific heat
in an approximate way.
We assume that the thermal excitations of kink-antikink pairs
at a temperature $T\equiv \beta ^{-1}$ are regulated by the
averaged antikink energy defined by
\begin{equation}
 \langle \epsilon_{\rm a}\rangle _{\beta} = \frac{\displaystyle{
 \sum_{k=1}^{l}  (\epsilon_0+\frac{1}{m}\cos\frac{\pi k}{l})
     \exp[-\beta (\epsilon_0+\frac{1}{m}\cos\frac{\pi k}{l})]
                                      }}
            {\displaystyle{
 \sum_{k=1}^{l}
     \exp[-\beta (\epsilon_0+\frac{1}{m}\cos\frac{\pi k}{l})]
 \label{eq:epsav}                                     }}.
\end  {equation}
Here the integer $l$ is the average distance between two excited kinks, i.e.,
the length of the region where an antikink can move.
We have neglected the effects of interactions between kinks and antikinks.
This may be justified if $l \gg 1$, which will be shown to be the case below.
We assume that the specific heat is written as a Schottky-type one as
\begin{equation}
 C(T) = \frac{(\beta\epsilon)^2 \exp[\beta\epsilon]}
              {(1+\exp[\beta\epsilon])^2},
   ~~~ \epsilon=\langle \epsilon_{\rm a}\rangle _{\beta}/2.
 \label{eq:c}
\end  {equation}
The average length $l$, which is equal to the distance between two
antikinks, is given by
\begin{equation}
l = {\rm int}(\exp[\beta \langle \epsilon_{\rm a}\rangle _{\beta}]),
\label{eq:l}
\end{equation}
since an antikink is excited with the probability
$\exp[-\beta \langle \epsilon_{\rm a}\rangle _{\beta}]$.
We can solve self-consistent equations (\ref{eq:epsav}) and (\ref{eq:l})
by iteration for a given $T$.
Then with obtained $\langle \epsilon_{\rm a}\rangle _{\beta}$,
we can calculate the specific heat at the temperature through Eq. (\ref{eq:c}).
We have employed the parameters given by Eqs. (\ref{eq:ep}) and (\ref{eq:mass})
and plotted the obtained specific heat in Fig. \ref{fig:spec}.
The peak position shifts from the finite-size result toward the low
temperature and locates at $T\sim 0.05$.
The profile of the peak is very steep on its low-temperature side
but it has a tail on the other side.
These features are common with the numerical results of finite-size systems.
This agreement implies that above approximation is
correctly taking into account of the temperature dependence of
$\langle \epsilon_a\rangle _{\beta}$.
The approximation is based on the idea that
an antikink can be treated as a localized object
after averaging the kinetic energy in
$\langle \epsilon_{\rm a}\rangle _{\beta}$.
We consider that this picture is a good approximation
if the discreteness of the antikink kinetic energy is negligible
and the motion of antikinks is well thermalized.
In fact, the average antikink energy $\langle\epsilon_{\rm a}\rangle_{\beta}$
obtained in the above calculation turns out to be $0.244$
at $T\sim0.05$, the peak position of the specific heat.
Therefore the averaged kinetic energy
$\langle\epsilon_{\rm a}\rangle_{\beta} - \epsilon_{0} = 0.029$
is larger than  $T/2=0.025$,
the classical thermal average of the kinetic energy.
The result implies that the phase
space of the $l$ states is well thermalized at this temperature.
The average distance between antikinks $l$ is estimated as $\sim 130$,
which is much greater than unity.
We find that the average kinetic energy is much smaller than the creation
energy at this temperature range .
Above results indicate that our approximate scheme
is a consistent one and is correctly
describing the thermodynamics of the system
in the temperature region of the peak in the specific heat.

\section{Conclusion}
\label{sec:conc}
In the previous sections, we have clarified the elementary excitations
in the $\Delta$-chain to a great extent.
They are two types of domain walls called a kink and an antikink.
A kink is essentially localized in a finite range (about one triangle).
An antikink propagates with kinetic energy within a region bounded by kinks.
The dispersionless mode found previously \cite{kubok93} originates
in the localized character of a kink.
The low-temperature peak in the specific heat
is caused by thermal excitations of kinks and antikinks.
The peak position is mainly determined
by the creation energy of the antikink,
while its kinetic energy causes the size-dependence
of the excitation energies as well as
the broadening of the peak of the specific heat.

Above understanding of the $\Delta$-chain might give some insight into the
properties of the {\it kagom\'e} antiferromagnet.
Though the dimensionality of two systems are different,
they have some common features.
Both systems have macroscopically degenerate classical
ground states and show the second peak in the specific heat at low
temperatures.
The lower peak in the specific heat is observed in finite {\it kagom\'e}
antiferromagnets.\cite{elser89,elstner-y94,zeng-e95,nakamura-m95}
The uniform susceptibility seems to decrease suddenly
at the same temperature.\cite{elstner-y94,nakamura-m95}
In the $\Delta$-chain, the spin gap corresponds to the first excitation gap.
Above similarities suggest that the elementary excitations with a spin gap
in both systems may have common characters that create the second peak of
the specific heat.
Recently, Zeng and Elser \cite{zeng-e95}
investigated the {\it kagom\'e} antiferromagnet by dimer calculations.
Their results support this speculation, but
further investigation on the {\it kagom\'e} antiferromagnet
is necessary for a concrete understanding of the system.

\acknowledgments

Numerical diagonalization was done with TITPACK Ver. 2. programed by
Professor H. Nishimori.
Use of the random number generator RNDTIK programmed by N. Ito and
Y. Kanada is gratefully acknowledged.
Computations were done partly on FACOM VPP500
at the Science Information Processing Center, University of Tsukuba,
and partly on FACOM VPP500 at the Institute for Solid State Physics,
University of Tokyo.

\centerline{Figure Captions}

\begin{figure}
 \caption{
   (a) A ground state of the periodic $\Delta$-chain with
       L-dimers, $\psi_{\rm g}^{\rm L}$, and
   (b) that with R-dimers, $\psi_{\rm g}^{\rm R}$.
       An ellipse stands for the dimer singlet pair.
   (c) A ground state of an open $\Delta$-chain.
   (d) An excited state of the periodic chain. The up spin
       that has a dimer singlet pair in its triangle
       is what we call a `kink' and the other one is an `antikink'.
   \label{fig:scheme}
         }
\end  {figure}
\begin{figure}
 \caption{(a) `Open-A' system. (b) `Open-B' system
 \label{fig:openscheme}
         }
\end  {figure}
\begin{figure}
  \caption{The nearest-neighbor spin correlations and the local
           magnetization in the Open-A system with 11 triangles (25 spins).
           Data for the lowest four states are shown.
  \label{fig:open-acorsz}
          }
\end  {figure}
\begin{figure}
  \caption{
           Domain wall profile,
           or the deviation from the dimer singlet state.
           Data of systems with 3, 7 and 11 triangles are plotted together.
  \label{fig:open-adom}
          }
\end  {figure}
\begin{figure}
 \caption{(a) A basis state for the 1-spin variation.
          (b) A basis state for the 5-spin variation.
 \label{fig:varbase}
         }
\end  {figure}
\begin{figure}
 \caption{(a), (b), (c)
          The wave function of an antikink for the lowest three levels
          obtained by the 5-spin variation.
          System size is $N=98$.
          (d)
          The excitation energy $\delta E$ is also plotted together with
          that by the 1-spin variation.
 \label{fig:varwave}
         }
\end  {figure}
\begin{figure}
 \caption{
          The local magnetization of Open-A system with 25 spins(11 triangles)
          evaluated numerically and variationally.
 \label{fig:varsiz}
         }
\end  {figure}
\begin{figure}
 \caption{Energy gaps of the lowest three excited states
          of the periodic chain in the $S=1$ subspace
          are plotted against $1/N^2$ together with those of the Open-A system
          obtained by numerical diagonalization and the variation.
 \label{fig:altogether}
         }
\end  {figure}
\begin{figure}
 \caption{The nearest-neighbor spin correlation and the local magnetization
          of the `Open-B' system that has a triplet state in the leftmost
          edge bond. The number of spins is 18.
          All of these three states are triplet.
 \label{fig:open-bedge}
         }
\end  {figure}
\begin{figure}
 \caption{
      (a)The nearest-neighbor correlation of the first excited state with $S=0$
         in the Open-B system and that with $S=1$.
         The system size is $N=9$, or the number of spin is 20.
      (b)The nearest-neighbor correlation of the ground state
         in the Open-A system
         and those of the first excited state in the Open-B system
         with $S=0$ and $S=1$.
         The number of spins are as denoted in the figure.
         The profile is depicted from the left for the $S=0$ state
         and the Open-A ground state, while its mirror image is
         depicted for the $S=1$ state.
 \label{fig:open-bs01}
         }
\end  {figure}
\begin{figure}
 \caption{The excitation energy of the periodic chains
          in the subspace with zero total momentum.
          The number of triangles are denoted by $N$.
 \label{fig:periodicgap}
         }
\end  {figure}
\begin{figure}
 \caption{Excitation energy of all the systems considered in this paper
          are plotted against $1-\cos \pi k/(N'+1)$.
          Definition of $N'=N+\Delta N$ is described in the text.
          The least-square fitting gives  $\epsilon_0=0.215$ and $m=1.21$.
 \label{fig:cosineplot}
         }
\end  {figure}
\begin{figure}
 \caption{Temperature dependence of the specific heat.
          Numerically exact results (solid line) and the
          approximate results (dashed line) for the $N=8$ periodic system
          are plotted together with the one estimated
          in the thermodynamic limit (bold line).
          Excitation energy in the Schottky-type specific heat
          $\epsilon=\langle\epsilon_{\rm a}\rangle_{\beta}/2$
          is also plotted.
 \label{fig:spec}
         }
\end  {figure}
\begin{table}[b]
\begin{center}
\begin{tabular}{lcc}
 System &~~~~&$\Delta N$ \\
\hline
 Open-A                & & 3              \\
 Open-B($S=0$)         & & 0.8            \\
 Open-B($S=1$)         & & 2.3            \\
 Periodic ($S=0, k=1$) & & -3.0           \\
 Periodic ($S=0, k=2$) & & -3.0           \\
 Periodic ($S=0, k=3$) & & -2.8           \\
 Periodic ($S=1, k=1$) & & -0.2           \\
 Periodic ($S=1, k=2$) & & -0.4           \\
 Periodic ($S=1, k=3$) & & -0.5           \\
\end{tabular}

\caption{$\Delta N$ for the systems we considered.}
\end{center}
\end{table}
\end{document}